\documentclass[twocolumn,showpacs,preprintnumbers,amssymb,eqsecnum]{revtex4}
\usepackage{latexsym,amsmath,graphicx}

\begin{document}
\title{Energy Extraction from Higher Dimensional Black Holes and Black Rings}

\author{Masato Nozawa$^{1}$}
\email{nozawa@gravity.phys.waseda.ac.jp}
\author{Kei-ichi Maeda$^{1,2,3}$}
\email{maeda@gravity.phys.waseda.ac.jp}
\address{$^{1}$Department of Physics, Waseda University, 3-4-1 Okubo,
Shinjuku-ku, Tokyo 169-8555, Japan~}
\address{$^2$ Advanced Research Institute for Science and Engineering,
Waseda  University, Shinjuku, Tokyo 169-8555, Japan~}
\address{$^3$ Waseda Institute for Astrophysics, Waseda University,
Shinjuku,  Tokyo 169-8555, Japan~}

\date{\today}

\begin{abstract}
We analyze the energy extraction by the Penrose process in higher dimensions. 
Our result shows the efficiency of the process from higher
dimensional black holes and black rings 
can be rather high compared with  than that in four dimensional Kerr
black hole. In particular, if one rotation parameter vanishes,
the maximum efficiency becomes infinitely large because the angular momentum 
is not bounded from above.
We also apply a catastrophe theory to 
analyze  the stability of black rings.
It indicates a branch of black rings with higher
rotational energy is unstable, which should be a different
 type of instability 
from the Gregory-Laflamme's one.
\end{abstract}
\pacs{04.50.+h, 04.70.Dy}
\maketitle

%========================Introduction============================%
\section{Introduction}
In recent years, a  brane world 
scenario has been extensively 
studied \cite{brane,large_extra_dimensions,brane_review}.
 One characteristic feature of brane world is
that mini black holes might emerge in the accelerator
\cite{collider}. If there exist large extra dimensions, 
the fundamental gravity scale $M_D$ could be order of TeV, much smaller
than $M_4\sim 10^{19}$GeV \cite{large_extra_dimensions}.
 Since the Schwarzschild radius of particles in
the collider becomes smaller than the particle radius, mini black holes
could be produced in the accelerators. If we are able to
detect the production and evaporation 
of black holes, it would be not only the direct
evidence for the Hawking radiation, but also the 
stepping stone to a unified theory.

For this reason, the basic study of higher dimensional black holes is
now of great importance. A naive estimate shows that mini-black holes can be
described by the classical solutions of the vacuum Einstein equations.
We may be able to ignore
the effects of brane.  In the colliders,
black holes would be
produced with rotation in general.

The rotating black holes in $D$ dimensional space-times with
$S^{D-2}$ topology were first derived by Myers and
Perry \cite{myers}. The black holes in higher dimensions
can have arbitrarily
large angular momentum unlike the case in 4-dimensional 
Kerr black hole.
The amazing discovery is the rotating black ring solution
by Emparan and Reall \cite{emparan}, whose topology of the event
horizon is $S^1 \times S^2$. 
Two black rings and one MP black hole with the
same mass and angular momentum can coexist. 
While in four dimensions, neutral charged black holes are
completely specified by their mass and angular momentum, and 
moreover non spherical topology is forbidden \cite{HE,uniqueness}. 
This is the first
counterexample that uniqueness theorem does not hold in higher
dimensions. It is then important to examine the property of the black
ring and black holes in detail. 

In this paper, we focus our attention on the energy extraction by the Penrose
process \cite{penrose,chandrasekhar,efficiency}.
Using such a formalism, we discuss the efficiency of the energy
extraction. 

The remainder of the paper is organized as follows: The
geometry of the higher dimensional black holes and black ring is summarized
in Section \ref{sec:hdbh}. Section \ref{sec:ee} consists of the main
discussions on the
energy extraction from higher dimensional black holes and black
rings. 
As for 5-dimensional objects,
adopting a catastrophe theory,
we analyze their stability in
Section \ref{sec:im}.
Our conclusions and some remarks follow 
in section \ref{sec:discussions}.

%=============================================================%
%
\section{Higher dimensional black holes and black rings}
\label{sec:hdbh}
%-------------------------even dimensions---------------------%
We first summarize black hole and black ring solutions in higher dimensions, 
which we discuss here.
\subsection{MP black hole solution}
Myers and Perry found an exact solution of the 
Einstein equations in arbitrary $D$ dimensional space-time \cite{myers}.
It represents a rotational black hole, which is a generalization of 
the four-dimensional Kerr black hole.
The solution is described by  a different form depending on whether
a space-time dimension  is even or odd.
Then we write the solutions in order.
\subsubsection{Even dimensions {\rm (}$D=2(d+1)${\rm )}}
The metric in even dimensions is given by;
\begin{align}
\label{eq:mpeven}
 ds^2 &= - dt^2+r^2 d\alpha ^2+\sum_{i=1}^d 
\left(r^2+a_{i}^2\right)\left(d\mu _{i}^2
+\mu_{i}^2d\phi  _{i}^2\right)\nonumber 
\\
\qquad &+
\frac{M r}{{\it \Pi}{\it F}}\left(dt+\sum_{i=1}^d a_i \mu_{i}^2 d\phi _{i}
\right)^2
+\frac{{\it \Pi} F}{{\it \Pi}-M r}dr^2,
\end{align}
where
\begin{eqnarray}
&& F = 1-\sum_{i=1}^d \frac{a_i ^2 \mu _i ^2}{r^2+a_i^2}, 
\label{eq:F}\\
&& {\it \Pi}=\prod_{i=1}^d (r^2+a_i^2),\label{eq:Pi}\\
&& \sum_{i=1}^d \mu _i ^2 +\alpha^2 = 1\,,
\end{eqnarray}
with $d\equiv D/2-1$.
%-----------------------M and J------------------------------%
The parameters $M$ and $a_i$
are related to the mass ${\cal M}$ and angular momenta ${\cal J}_i$
as
\begin{eqnarray}
 && {\cal M}=\frac{D-2}{16\pi G}A_{(D-2)}M,\\
 && {\cal J}_i =\frac{1}{8\pi G}A_{(D-2)}M a_i ~~~~~~\, (i=1, \cdots, d)\,,
\end{eqnarray}
where $G$ and $A_{(D-2)}$ are $D$-dimensional gravitational constant and 
 the area of a unit ($D-2$)-sphere, which is given by
\begin{eqnarray}
 A_{(D-2)}=\frac{2\pi ^{(D-1)/2}}{\Gamma ((D-1)/2)}\,,
\end{eqnarray}
respectively.

The event horizon appears where 
\begin{eqnarray}
g^{rr}={{\it \Pi} -Mr\over {\it \Pi} F}
\end{eqnarray}
 vanishes. 
If at least one rotation parameter set to zero, for example $a_1=0$, 
the equation for  horizon is given by 
\begin{eqnarray}
 {\it \Pi}-Mr=r^2 \left(\prod_{i\geq 2}^d (r^2+a_i^2)-\frac{M}{r}\right)=0.
\label{eq:horizon}
\end{eqnarray}
In the case of $d\geq 2$,  i.e.  $D\geq 6$,
Eq. (\ref{eq:horizon}) has a positive 
root independent of
the magnitude of $a_i$. We then find a regular black hole solution 
albeit the angular momenta are
arbitrarily large. This is one of typical features of higher dimensional
black holes.

%-----------------------odd dimensions---------------------%
\subsubsection{Odd dimensions {\rm (}$D=2d+1${\rm )}}
 In odd dimensions, the metric of a rotating black hole is slightly
changed from Eq. (\ref{eq:mpeven}), which is given by 
\begin{align}
\label{eq:mpodd}
ds^2 &= - dt^2+\sum_{i=1}^d\left(r^2+a_{i}^2\right)
\left(d\mu _{i}^2+\mu _{i}^2 d\phi _{i}^2\right)
\nonumber\\
\qquad &+
\frac{M
 r^2}{{\it \Pi}{\it F}}\left(dt+\sum _{i=1}^d a_i \mu_{i}^2 d\phi _{i}\right)^2
+\frac{{\it \Pi F}}{{\it \Pi}-M r^2}dr^2 ,
\end{align}
with
\begin{eqnarray}
\sum _{i=1}^d \mu _i ^2=1 ,
\end{eqnarray}
where the definition of $\it{\Pi}$ and 
$d=(D-1)/2$ . 
We also find that if at least two angular momenta set to zero, 
the remaining angular
momenta can be arbitrarily large for $d\geq 3$,  i.e.  $D\geq 7$
 as in the case of even dimensions,
because the equation for horizon is now
\begin{eqnarray}
 {\it \Pi}-Mr^2=r^4 \left(\prod_{i\geq 3}^d 
(r^2+a_i^2)-\frac{M}{r^2}\right)=0\,.
\label{eq:horizon_odd}
\end{eqnarray}

%-----------------------MP in five dimensions---------------------%
A five-dimensional black hole is exceptional, because there is 
an upper bound for the angular momentum.
We also write down a five-dimensional 
black hole solution 
with two rotation parameters $a$ and $b$
 in the Boyer-Lindquist coordinates, which  
is given by 
\begin{align}
\label{eq:mp5}
 ds^2 =& -dt^2 +\frac{\rho^2 r^2}{\Delta}dr^2+\rho^2
  d\theta  ^2 \nonumber \\
&+\frac{ M }{\rho ^2}(dt+a\sin ^2 \theta d\phi +b\cos ^2 \theta
 d\psi)^2  \\
& +(r^2+a^2)\sin ^2 \theta d\phi ^2+(r^2+b^2)\cos ^2 \theta d\psi ^2 ,\nonumber
\end{align}
where
\begin{eqnarray}
&& \rho ^2=r^2+a^2\cos ^2\theta +b^2\sin^2\theta ,\\
&& \Delta =(r^2+a^2)(r^2+b^2)-M r^2 .
\end{eqnarray}
The horizon appears where $\Delta =0$, which gives the location of 
the horizons, i.e.  
\begin{eqnarray}
r_{\pm}^2 
\equiv \frac{1}{2}\left[
M -(a^2 +b^2)\pm \sqrt{[M -(a+b)^2][M-(a-b)^2]}\right].
\nonumber 
\\
&~&
\end{eqnarray}
The sign change of rotation parameters $a, b$ 
simply reverses the direction of rotation.
The condition for the existence of an event horizon is
\begin{eqnarray}
\label{eq:above5mp}
M \geq (|a|+|b|)^2.
\end{eqnarray}
The outer and inner horizons coincide when $M =(|a|+|b|)^2$. 
The area of the event horizon is given by 
\begin{eqnarray}
 {\cal A}_{H}={2\pi ^2 \over r_{+}}(r_+^2+a^2)(r_+^2+b^2).
\end{eqnarray}
The horizon vanishes if one of the angular parameters set to zero
and the other approaches the extreme value  (e.g. $b=0$ and $a^2 \to M$), 
which corresponds to the appearance of a naked singularity.
When $(|a|+|b|)^2 \rightarrow M $ with $a\neq0$ and $b\neq0$, 
this corresponds to the extremal black hole with non-zero surface 
area and vanishing temperature.

%------------------------------------------------------------%
%----------------------black ring----------------------------%
\subsection{a black ring solution}
Emparan and Reall found a new exact solution of the vacuum 
Einstein equations in five
dimensions, which is asymptotically flat, stationary and regular on and
outside the event horizon \cite{emparan}. This solution describes a black ``hole''
 with the ring topology $S^1 \times S^2$, which is called a black ring. 

%---------------------metric------------------------------%
The metric of a rotating black ring is written in the following form
\cite{emparan2,cmetric}
\begin{align}
\label{eq:blackring}
 ds^2 =& -\frac{F(x)}{F(y)}\left(dt+R\frac{\sqrt{\lambda \nu
 (1+\lambda)}}{1+\nu}(1+y)d\psi\right)^2
 \\
& +\frac{R^2}{(x-y)^2}\left[-F(x)\left(\frac{(1+\lambda)G(y) }{(1+\nu
 )^2}d\psi  ^2+\frac{F(y)}{G(y)}dy^2
 \right)\right. \nonumber \\
&\left.
 \hspace{2cm}+F(y)^2\left(\frac{dx^2}{G(x)}+\frac{G(x)(1+\lambda
 )}{F(x)(1+\nu)^2}d\phi ^2 
 \right)\right].\nonumber
\end{align}
where
\begin{eqnarray}
 && F(\xi )=1-\lambda \xi ,\\
 && G(\xi )=(1-\xi ^2)(1-\nu \xi ). 
\end{eqnarray}
Here we have adopted a  form of the C-metric introduced in \cite{cmetric}.

%-------------------------parameter range------------------------------%
The parameter $R (>0)$ is interpreted a ``radius'' of the ring. The
parameter $\nu $, related to a ``thickness'' of the ring \cite{emparan2},
takes the value in $0<\nu <1$. Both angular coordinates
$\phi $ and $\psi $  have period $2\pi$. $x, y$ take the value in the range
\begin{eqnarray}
 -1\leq x \leq 1, \qquad -1 <y^{-1} <\lambda.
\end{eqnarray}
$x=+1$ corresponds to the inside equatorial plane of the ring, while $x=-1$ 
denotes  the outside of the equatorial plane.
The spatial infinity corresponds to  $x\neq -1$ and $y\rightarrow -1$, 
while the symmetric axis is given by $x=+1$ and $y\rightarrow -1$.
A space-time singularity
is located at $y=\lambda^{-1}$. The event
horizon and the ergosurface are  given by 
$y=\nu ^{-1}$ and $y^{-1}=0$, respectively. 
%------------------------topology--------------------------------%
We find  a regular black ring solution if we impose 
some relation between $\lambda$ and $\nu$.
This metric form also includes a black hole solution for 
a different relation of  $\lambda$ and $\nu$.
Then the parameter $\lambda $ determines the topology of the objects,
i.e.,
\begin{eqnarray}
 \lambda =\left\{
\begin{array}{ll}
\quad 1\qquad &(\textrm{a black hole})\\ \label{eq:lambda}\\
\displaystyle \frac{2\nu }{1+\nu ^2} \qquad & (\textrm{a black ring}).
\end{array}
\right.
\end{eqnarray}
%--------------------- coordinate transformation -----------------%
The  solution of $\lambda =1$ corresponds to the five-dimensional MP
 black hole with one rotation
parameter. In fact, Eq. (\ref{eq:blackring}) is transformed into 
Eq. (\ref{eq:mp5}) with $b=0$, by the coordinate transformation
\cite{emparan}
\begin{align}
r^2= M\frac{(1-y)(1-\nu x)}{(1+\nu )(x-y)},\quad
\cos ^2\theta =\frac{(1-y)(1+x)}{2(x-y)},
\end{align}
with
\begin{align}
M =\frac{4R^2}{1+\nu},\qquad
a=\frac{2\sqrt{2\nu }R}{1+\nu }.
\end{align}
%------------------------physical parameters----------------------%
The mass, angular momentum, surface gravity, angular velocity and
horizon area are given by \cite{emparan2}
\begin{align}
\label{eq:brmass}& {\cal M}=\frac{3\pi R^2}{4G}\frac{\lambda (\lambda +1)}{\nu
 +1}, \enspace
 {\cal J}=\frac{\pi R^3}{2G}\frac{\sqrt{\lambda \nu }(\lambda
 +1)^{5/2}}{(1+\nu)^2}.\\
\label{eq:brgravity}& \kappa =\frac{1}{2R}\frac{1-\nu }{\lambda
 ^{1/2}(\lambda -\nu 
 )^{1/2}}, \quad
\Omega _H=\frac{1}{R}\sqrt{\frac{\nu }{\lambda (1+\lambda )}},\\
\label{eq:brarea}& {\cal A}_{H}=8\pi ^2 R^3 \frac{\lambda ^{1/2}(1+\lambda
 )(\lambda -\nu 
 )^{3/2}}{(1+\nu )^2(1-\nu )}.
\end{align}
The horizon area vanishes as $\nu \to 1$, and in this
limit, a naked singularity appears. 
The opposite limit of $\nu \to 0$ leads a 
solution without an angular momentum. 
This is a static black ring solution with
a conical singularity \cite{weyl}.

%----------------------nonunique----------------------------%
We introduce a dimensionless reduced spin parameter $j$ by
\begin{eqnarray}
 j^2=\frac{27\pi}{32G}\frac{{\cal J}^2}{{\cal M}^3}.
\end{eqnarray}
Using Eqs. (\ref{eq:lambda}) and (\ref{eq:brmass}), we find
\begin{eqnarray}
 j^2 =\left\{
\begin{array}{ll}
\displaystyle \frac{2\nu }{1+\nu }\quad &(\textrm{a black hole})\\\\
\displaystyle \frac{(1+\nu )^3}{8\nu } \quad &(\textrm{a black ring}).
\end{array}
\right.
\end{eqnarray}

%---------------------fig.nonunique----------------------%
%\begin{figure}[h]
%\begin{center}
%\includegraphics[width=7cm]{nonunique.eps}
%\caption{The solid line represents a black ring, the thin line a MP
% black hole. There exist two black rings and a black hole for the  
% same value of $j$ in the range of
% $27/32<j^2<1$.}
%\label{nonunique}
%\end{center}
%\end{figure}
%--------------------------------------------------------%
There exist two black rings and a black hole with the same mass and spin in
the range of $27/32<j^2<1$. Two rings are distinguished by their area
(entropy) \cite{emparan}.

%==================================================================%
%=======================  Energy extraction  ======================%
%==================================================================%

\section{Energy extraction}
\label{sec:ee}

%------------------------even------------------------------%

In this section, we discuss the Penrose procees
\cite{penrose,chandrasekhar,efficiency}, by which we can extract a rotation energy
from a black hole or a black ring.
In the Penrose process, an incident particle 
with the $D$-momentum $p_{(0)}^\mu$
is supposed to split into
two fragments (first and second particles with the $D$-momenta 
$p_{(1)}^\mu$ and 
$p_{(2)}^\mu$)
in the ergoregion. 
One of them crosses the event horizon, 
while the other escapes to infinity.

In order to calculate the efficiency of 
energy extraction by
the Penrose process, we consider a very simple case, that is, all
particles are confined on one plane, which we call  an ``equatorial'' plane.
In higher dimensions, there are several planes on which a particle's 
trajectory can be confined. 
%---------------------------conservation----------------------------%
At the point of split, the total $D$-momentum is conserved as
\begin{eqnarray} 
p_{(0)}^\mu = p_{(1)}^\mu+p_{(2)}^\mu\,. 
\end{eqnarray} 
The momenta of three 
particles $p^{\mu }_{(I)}$ ($I=0, 1,2$) 
are nonspacelike and hence should lie inside a local light cone.

The orbit of the particle moving on a plane
is described by 
two dimensional coordinates, i.e. 
radial and one angular coordinates ($r$  and $\phi$).
Then we can write  the momentum of a particle along the geodesic 
$\gamma $ as
%---------------------------velocity--------------------------------%
\begin{equation}
p_\gamma  
=p^\mu{\partial \over \partial x^\mu}=
p^t
\left(\frac{\partial }{\partial t}+v\frac{\partial }
{\partial r}+ \Omega\frac{\partial }{\partial \phi}
\right),
\end{equation}
with
\begin{eqnarray}
v=\frac{dr}{dt}, \qquad \Omega
 =\frac{d\phi}{dt}\,.
\end{eqnarray}
The relation $E=-p_{t}$ yields
\begin{align}
p^t=-\frac{E}{X}, \qquad X\equiv g_{tt}+\Omega g_{t\phi}.
\label{eq:X1}
\end{align}
From $p^{\mu }p_{\mu }=-m^2$,
we find 
\begin{eqnarray}
g_{tt}+2g_{t\phi}\Omega+g_{\phi \phi}
\Omega ^{2}
=-g_{rr}v^{2}-\left({mX\over E}\right)^{2}\leq 0\,.
\label{eq:X2}
\end{eqnarray}
%--------------------------null angular velocity-------------------------%
Then the angular velocity with respect to an asymptotic infinity observer 
($\Omega$)
takes the
value in the range of $\Omega ^{-}\leq \Omega \leq \Omega ^{+}$, where
\begin{equation}
 \Omega^{\pm}\equiv \omega\pm \sqrt{\omega^{2}-\frac{g_{tt}}
{g_{\phi \phi}}},
\end{equation} 
with
%-------------------------angular velocity of LNRO-----------------------%
\begin{equation}
\omega=-\frac{g_{t\phi}}{g_{\phi  \phi }}\,,
\end{equation}
which denotes the angular velocity of a locally nonrotating observer at a
given radius $r$.
The conservation of energy ($E=-p^t X$) and angular momentum ($L=p^t \Omega $)
are written as
\begin{eqnarray}
p^t_{(0)} X_{(0)}&=&p^t_{(1)} X_{(1)}+p^t_{(2)} X_{(2)},
\label{energy_cons}\\
p^t_{(0)}\Omega _{(0)}&=&p^t_{(1)}\Omega _{(1)}+p^t_{(2)}
\Omega_{(2)}.
\label{angular_cons}\end{eqnarray}
%------------------------  efficiency  -------------------------------%
Suppose that the first particle crosses the horizon with negative energy
$E_{(1)}<0$.
The second particle will get its energy when it escapes into infinity.
The efficiency of the Penrose process is then given by
\begin{eqnarray}
 \eta =\frac{E_{(2)}-E_{(0)}}{E_{(0)}}=\chi-1 ,
\end{eqnarray}
From the energy and angular momentum conservation 
(Eqs. (\ref{energy_cons}) and 
(\ref{angular_cons}))
with the definition of $X$ (Eq. (\ref{eq:X1})), 
\begin{align}
 \chi &= \frac{E_{(2)}}{E_{(0)}}
=\frac{(\Omega _{(0)}X_{(1)}-\Omega _{(1)}X_{(0)})X_{(2)}}
{(\Omega _{(2)}X_{(1)}-\Omega _{(1)}X_{(2)})X_{(0)}}\nonumber \\
&=
\frac{(\Omega _{(0)}-\Omega _{(1)})X_{(2)}}{(\Omega _{(2)}-
\Omega _{(1)})X_{(0)}}.
\end{align}
Here we consider the case that the incident particle has
zero initial velocity, i.e. $E_{(0)}=m_{(0)}$, and assume that
it will decay into two photons \cite{chandrasekhar}, 
i.e. $p_{(1)}$ and $p_{(2)}$
are null.
The efficiency $\eta =\chi -1$ is maximized 
if we have the largest value of $\Omega _{(2)}$
and the smallest one of $\Omega _{(1)}$ simultaneously, 
which 
is obtained when all $v_{(I)}$ vanish.
In that case, we find 
\begin{align}
&p_{(1)}=p^t_{(1)}\left(\frac{\partial }{\partial t}+
\Omega ^-
\frac{\partial }{\partial \phi}\right),\\
&
p_{(2)}=p^t_{(2)}\left(\frac{\partial }{\partial t}+
\Omega ^+
\frac{\partial }{\partial \phi}\right).
\end{align}
with
%-------------------------- Omega max --------------------------------%
\begin{eqnarray}
\Omega _{(1)}=\Omega ^{-}, \qquad
 \Omega _{(2)}=\Omega ^{+},
\end{eqnarray}
and from Eqs. (\ref{eq:X1}) and (\ref{eq:X2}),
\begin{align}
\Omega_{(0)}=\frac{-g_{t\phi}(1+g_{tt})+\sqrt{(1+g_{tt})
(g_{t\phi}^2-g_{tt}g_{\phi\phi})}}
{g_{\phi \phi}+g_{t\phi}^2}\,.
\end{align}
%
%========================= Fig lightcone =====================%
\begin{figure}[h]                                             
\begin{center}
\includegraphics[width=5cm]{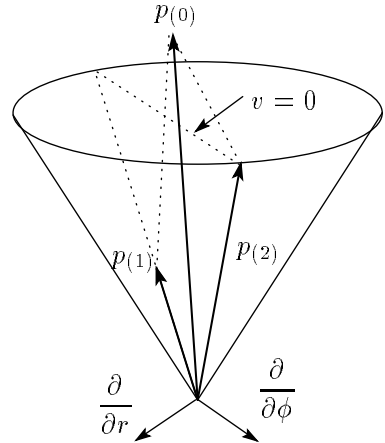}
\end{center}
\caption{The momenta $p_{(I)}$ lie in the local light cone. 
The maximum efficiency of the Penrose process
is achieved when the radial velocities 
$v_{(I)}$ vanish,
assuming $p_{(1)}$ and $p_{(2)}$
are null.}
\label{fig:lightcone}
\end{figure}\\
%=============================================================%
The maximum efficiency is then given by
%-----------------------Maximal efficiency------------------------%
\begin{align}
\eta _{\rm max}&=\chi _{\rm max}-1 
={(\Omega _{(0)}-\Omega ^{-})(g_{tt}+g_{t\phi}\Omega ^{+})\over
(\Omega ^{+}-\Omega ^{-})(g_{tt}+g_{t\phi}\Omega _{(0)})} -1\nonumber \\
\nonumber \\
&\leq {g_{\phi \phi}(\sqrt{1+g_{tt}}+1)+g_{t\phi}^2\over
2g_{\phi \phi }\sqrt{1+g_{tt}}}-1,
\label{eq:efficiency}
\end{align}
where the equality holds when the split occurs at the horizon. 
%-------------------------Kerr------------------------------%

In four dimensions, the rotation parameter of Kerr 
black hole has an upper bound
$a\leq {\cal M}$, then we find
\begin{align}
\eta _{\rm max}&= \chi _{\rm max}-1
 =\frac{1}{2}\left(\sqrt{\frac{2{\cal M}}{r_+}}-1\right)\nonumber \\
& \leq 
\frac{1}{2}(\sqrt{2}-1)\sim 20.7\%.
\label{eq:max}
\end{align}
The equality holds when the split occurs at the event horizon and 
the black hole is extreme ($a={\cal M}$).
Then we find the maximum efficiency of the Penrose process 
in the Kerr space-time is about
$20.7\%$. This recovers the results in 
\cite{chandrasekhar,efficiency}.

%-------------------------------------------------------------------%
Now we discuss the efficiency of the Penrose process in higher dimensions
for each solution in order.

\subsection{Black Hole}
\subsubsection{Even dimensions {\rm (}$D=2(d+1)${\rm )}}

%==========================efficiency================================%
%--------------------------------------------------------------------%
 The coordinates $\mu _i$ and $\alpha $ in the metric (\ref{eq:mpeven}),
are written explicitly by
 colatitude angle $\theta_i$ as follows:
\begin{eqnarray}
 \left\{
\begin{array}{ll}
& \mu_1=\sin \theta _1\\
& \mu_2=\cos \theta_1\sin \theta_2  \\
& \quad \vdots  \\
& \mu_d=\cos \theta _1\cos \theta _2 \cdots \sin \theta_d  \\
& \alpha =\cos \theta _1 \cos \theta _2 \cdots \cos \theta_d\,.
\end{array}
\right.
\end{eqnarray}
Suppose that  the orbits of particles are constrained on the
 ``equatorial'' plane
$\theta_1=\theta_2=\cdots=\theta_d=\pi /2$.
Note that since each coordinate $\mu_i$ is an equal footing, 
we can exchange the numbering of $\mu_i$. 
Then we find $d$ ``equatrial'' planes,
on which the orbits of particles are confined.
As a result, we can replace $a_1$ in the result obtained 
below with any other rotation parameter $a_i$.

%-------------------------------------------------------------------%
Substituting the metric (\ref{eq:mpeven}) at the horizon $r_+$ 
into Eq. (\ref{eq:efficiency}), 
we find the maximum efficiency $\eta_{\rm max}$ as
\begin{eqnarray}
 \eta \leq \eta_{\rm max}\equiv \frac{1}{2}\left(\sqrt{1+\frac{a_1^2}{r_+^2}}-1 
\right)\,.
\label{eq:effeven}
\end{eqnarray}
The black hole event horizon $r_+$ is given by 
\begin{eqnarray}
 \prod _i(r_+^2+a_i^2)-M r_+=0.
\end{eqnarray}
In six or higher dimensions, the angular parameter $a_1$ can take
arbitrarily large value if at least one remaining rotation parameter 
is zero. Thus in this case,
 we are able to extract quite a lot of energy from black holes.

%=========================== D=6 case ==============================%
We shall see this fact more explicitly.
For simplicity, we first consider the case of $D=6$.
In this case, we have two rotation parameter.
We set $a_1=a, a_2=\beta a$ $(a>0, \eta>0)$. $\beta$ denotes the ratio of 
the second rotation parameter to the first one.
Then the equation for the horizon leads
\begin{eqnarray}
(x_+^2+1)(x_+^2+\beta^2)-\tilde{M} x_+=0\,,
\end{eqnarray}
where $x_+=r_+/a, \tilde{M}=M/a^3$.
The horizon is larger than the critical value
\begin{eqnarray}
x_{\rm cr}^2\equiv {1\over 6}\left[-(1+\beta^2)
+\sqrt{(1+\beta^2)^2+12\beta^2}\right]\,,
\end{eqnarray}
The rotation parameter is limited, when $\beta$ is fixed, as
\begin{eqnarray}
a^3\leq {x_{\rm cr}\over (x_{\rm cr}^2+1)(x_{\rm cr}^2+\beta^2)}
M\,.
\end{eqnarray}
From Eq. (\ref{eq:effeven}), we obtain the maximum efficiency 
in terms of $\beta$ as
\begin{eqnarray}
\eta_{\rm max}&=&{1\over 2}\left(\sqrt{1+{1\over x_{\rm cr}^2}}-1\right)
\nonumber \\
&=&{1\over 2}\left(\sqrt{{1\over 2\beta^2}\left(1+3\beta^2
+\sqrt{(1+\beta^2)^2+12\beta^2}
\right)}-1\right)
\nonumber \\
~\,.
\end{eqnarray}
We summarize some typical cases in Table 1.
%
%====================== Table D=6 =======================%
\begin{table}[h]
\caption{The maximum efficiency $\eta_{\rm max}$
of the Penrose process in  MP black hole
with $D=6$. $a_1$ and $a_2$ are two rotation parameters.
$\beta=a_2/a_1$ and $r_+$ is the horizon radius.}
\begin{center}
\begin{tabular}{|c||c|c|c|}
\hline
&$a_1\ll a_2$&$a_1=a_2$&$a_1\gg a_2$ \\
\hline
\hline
$\beta$&$\rightarrow \infty$&$1$&$\rightarrow 0$ \\
\hline
$r_+$& $r_+>a_1$ & $r_+
>\dfrac{1}{\sqrt{3}}a_1$ &$r_+>\beta a_1$ \\
\hline
$a_1$&$a_1\leq \left(\dfrac{M}{2\beta^2}\right)^{1/3} $&
$a_1\leq \left(\dfrac{3\sqrt{3}M}{16}\right)^{1/3}$&
$a_1\leq \left(\dfrac{M}{2\beta}\right)^{1/3} $\\
\hline
$\eta_{\rm max}$&$\dfrac{1}{2}(\sqrt{2}-1)$&
$\dfrac{1}{2}$&$\dfrac{1}{2\beta} (\rightarrow \infty)$ \\
\hline
\end{tabular}
\label{table_1}
\end{center}
\end{table}
%===========================================================%
We find the maximum efficiency diverges as $1/(2\beta)$ when
the second rotation parameter $a_2$ decreases to zero, i.e.
$a_2=\beta a_1\rightarrow 0 ( \beta \rightarrow 0 )$.
On the other hand, even if the first rotation parameter $a_1$
is very small as $a_1=a_2/\beta\rightarrow 0 ( \beta \rightarrow \infty )$,
the efficiency does not vanish. It gives the same efficiency 
as that in four dimensional black hole.  Although this value is not so large,
we obtain the large efficiency if $a_2$ is large enough by putting the particles on
the different equatorial plane, i.e. $(\mu_2,\phi_2)$-plane.
This result is also obtained just by
the exchange of coordinates $(\mu_i,\phi_i)$
($i=1,2$).
As the result, if one rotation parameter is enough large,
we can extract rotational energy by any amount by the Penrose process. 

One interesting observation is the case of the same rotation parameters
; $a_1=a_2=a (\beta =1)$.
In this case, the horizon radius is limited as $r_+>a/\sqrt{3}$,
The efficiency is finite, i.e. $\eta_{\rm max}=1/2$.

%---------------------------equal spin case-----------------------%
This is true for any even dimensions.
If all rotation parameters are the same, i.e. 
$a_i=a (i=1, \cdots, d)$,
we find the maximum efficiency
by
\begin{eqnarray}
\eta_{\rm max}={1\over 2}\left(\sqrt{2d}-1 \right)\,.
\end{eqnarray}
In this case the horizon and rotation parameters are limited as
\begin{eqnarray}
r_+&>&{a\over \sqrt{2d-1}}\\
a&<&\left({(2d-1)^{d-1/2}\over (2d)^d }\right)^{1/(2d-1)}M^{1/(2d-1)}
\,.
\end{eqnarray}
The efficiency becomes larger as $d$ increases,
but not so much.

\subsubsection{Odd dimensions {\rm (}$D=2d+1${\rm )}}
%---------------------------odd----------------------------%
The efficiency is obtained by almost the same fashion as that in even
dimensions. The coordinates $\mu_i$ are explicitly written such that
\begin{eqnarray}
 \left\{
\begin{array}{ll}
& \mu_1=\sin \theta _1 \\
& \mu_2=\cos \theta_1\sin \theta_2  \\
& \quad \vdots  \\
& \mu_{d-1}=\cos \theta _1\cos \theta _2 \cdots \sin 
\theta_{d-1}  \\
& \mu_{d}=\cos \theta _1 \cos \theta _2 \cdots \cos 
\theta_{d-1} .
\end{array}
\right.
\end{eqnarray}
If the particle orbit is constrained on the ``equatorial'' plane 
($\theta_1=\theta_2=\cdots=\theta_{\frac{d-3}{2}}=\pi /2$)
as in five dimensions \cite{frolov}, 
we obtain the
efficiency of the Penrose process from Eqs. (\ref{eq:efficiency}) and 
(\ref{eq:mpodd}),
\begin{eqnarray}
 \eta \leq \frac{1}{2}\left(\sqrt{1+ \frac{a_1^2}{r_+^2}}-1\right),
\end{eqnarray}
where the  horizon radius $r_+ $ is given by
\begin{align}
\prod _i (r_+^2+a_i^2)-M r_+^2=0.
\end{align}
As in the case of even dimensions, 
$a_1$ can be arbitrarily large if $D\geq 7$ 
and at least two rotation parameters vanish, we could
have more productive energy extraction from a higher dimensional black hole
than from four dimensional Kerr black hole. 
However, if all rotation parameters are the same as
$a_i=a (i=1, \cdots, d)$,
we find the maximum efficiency
by
\begin{eqnarray}
\eta_{\rm max}={1\over 2}\left(\sqrt{d}-1 \right)\,.
\end{eqnarray}
In this case the horizon and rotation parameters are limited as
\begin{eqnarray}
r_+&>&{a\over \sqrt{d-1}}\\
a&<&\left({(d-1)^{d-1}\over d^d }\right)^{1/(2(d-1))}M^{1/(2(d-1))}
\,.
\end{eqnarray}
The efficiency becomes larger as $d$ increases,
but not so much.

We shall give the detail in five dimensions
with two rotation parameters $a=a_1$ and $b=a_2$.
The equation for the horizon is now
\begin{eqnarray}
(x_+^2+1)(x_+^2+\beta^2)-\tilde{M} x_+^2=0\,,
\end{eqnarray}
where $x_+=r_+/a$, $\beta=b/a$, and $\tilde{M}=M/a^2$.
We find that the horizon radius is limited as $x_+\geq \beta ^{1/2}$.
Then we obtain the maximum efficiency as
\begin{eqnarray}
\eta_{\rm max}={1\over 2}\left({\sqrt{1+\beta\over \beta}}-1
\right)
\,,
\end{eqnarray}
This result is summarized in Table 2.
%
%===================== Table D=5 ==========================%
\begin{table}[h]
\caption{The maximum efficiency $\eta_{\rm max}$
of the Penrose process in  MP black hole
with $D=5$. $a$ and $b$ are two rotation parameters.
$\beta=b/a$ and $
r_+$ is the horizon radius.}
\begin{center}
\begin{tabular}{|c||c|c|c|}
\hline
&$a\ll b$&$a=b$&$a\gg b$ \\
\hline
\hline
$\beta$&$\rightarrow \infty$&$1$&$\rightarrow 0$ \\
\hline
$r_+$& $r_+>\sqrt{\beta} a $ & $r_+>a$ &$r_+>\sqrt{\beta}
 a $ \\
\hline
$a$&$\displaystyle a\leq {\sqrt{M}\over \beta} $&
$\displaystyle a\leq {\frac{\sqrt{M}}{2}}$&
$a\leq \sqrt{M} $\\
\hline
$\eta_{\rm max}$&$\displaystyle {1\over 4\beta} (\rightarrow 0)$
&$\displaystyle {1\over 2}\left(\sqrt{2}-1\right)$
&$\displaystyle {1\over 2\sqrt{\beta}} (\rightarrow \infty)$ \\
\hline
\end{tabular}
\label{table_2}
\end{center}
\end{table}
%================================================================%

%================================================================%
%-------------------------- ring --------------------------------%
%================================================================%
\subsection{Black Ring}
Since infinity is at $x=y=-1$, we consider to
inject a particle from $x=-1$ direction. Writing down the geodesic
equation, we can see the particle is constrained $x=-1$ plane through
the orbit. In Appendix \ref{sec:app1}, we summarize a motion of 
a particle on the equatorial plane in a black ring spacetime. 

\begin{figure}[h]
\begin{center}
\includegraphics[width=7cm]{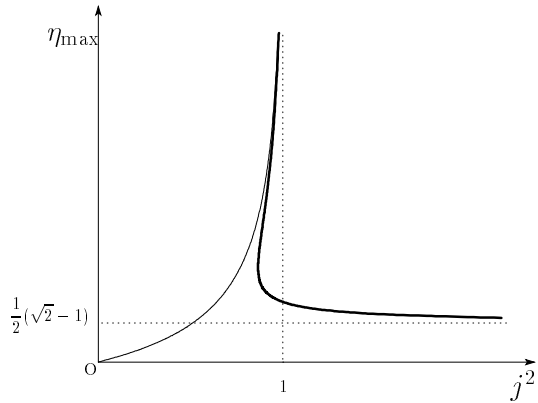}
\caption{The energy extraction efficiency in terms of a reduced spin.
The solid line corresponds to a black ring and the fine line to
a black hole.}
\label{fig:efficiency}
\end{center}
\end{figure}

The maximum efficiency of the process is given from Eq. (\ref{eq:efficiency}) 
as
\begin{eqnarray}
 \eta_{\rm max} 
&=& 
{1\over 2}
\left(\sqrt{1+\frac{\nu (1+\lambda )}{\lambda -\nu
}}-1\right)
\nonumber \\
&=&\left\{
\begin{array}{ll}
\displaystyle {1\over 2}\left(\sqrt{\frac{1+\nu}{1-\nu }}-1\right)
\qquad &(\textrm{a black hole})\\ \label{eq:brefficiency}\\
\displaystyle {1\over 2}\left(\sqrt{\frac{2}{1-\nu }}-1\right)
 \qquad & (\textrm{a black ring}).
\end{array}
\right.
\end{eqnarray}
Since $\nu $ takes value in the range $0< \nu < 1$, 
the efficiency is greater than that from 4-dimensional Kerr black hole.
This black hole solution corresponds to
MP black hole in five dimensions with one rotation parameter $a$.
In the limit of $\nu\rightarrow 1$, we find maximum efficiency 
($\eta_{\rm max}\rightarrow \infty$).
Compared a black ring with a black hole, the efficiency of a black ring
is larger than that of a black hole for the same value of $\nu$.
However, if we evaluate them for the same reduced angular momentum,
i.e. $j$, we find the opposite result (see Fig. \ref{fig:efficiency})
in the common range of $27/32<j^2<1$. 
From Fig.\ref{fig:efficiency} , 
we also find an interesting feature for a black ring.
The efficiency for larger angular momentum ($j^2>1)$
decreases with respect to $j$ to some finite constant.
The larger angular momentum does not provide the larger efficiency.
We will discuss its reason later. 

%---------------------Fig Efficiency-------------------------%
%-----------------------------------------------------------%

%================================================================%
%======================= irreducible mass =======================%
%================================================================%
\section{Irreducible mass and Rotational Energy}
\label{sec:im}

The relationship between the black hole mechanics and thermodynamics is
now well established \cite{bhtd}. The integrated mass formula holds 
for higher dimensional MP black holes and also black rings 
\cite{myers,emparan}.
\begin{eqnarray}
\label{eq:second}&& 
{\cal M}=\frac{d-2}{d-3}\left(
\frac{\kappa }{8\pi G}{\cal A}+\Omega _H{\cal J}\right).
\end{eqnarray}
The equation (\ref{eq:second}) in Kerr space-time is first
found by Smarr \cite{smarr}.
The mass of the neutral black hole consists of the surface
energy and the rotational energy \cite{smarr},
\begin{eqnarray}
 && {\cal E}_S\equiv \frac{1}{8\pi G}\int_0^{\cal A} \kappa({\cal A},{\cal J}=0)
d{\cal A} ,\\
 && {\cal E}_R\equiv \int_0^{\cal J} \Omega_H({\cal A},{\cal J})d{\cal J}
.\label{eq:irr}
\end{eqnarray}
The surface energy ${\cal E}_S$ is often referred as the irreducible mass
${\cal M}_{\rm{irr}}$, which cannot be decreased by any classical process. 
For the Kerr black hole, the irreducible mass is
\begin{align}
 {\cal M}_{\rm irr}
=\frac{1}{2}(r_+^2+a^2)^{1/2}.
\end{align}
It takes maximal value ${\cal M}/\sqrt{2}$ when the solution is extremal
$a={\cal M}$. Hence the rotational energy which we can extract at most is 
\begin{eqnarray}
 {\cal E}_R/{\cal M}
=\left(1-\frac{1}{\sqrt{2}}\right)\sim 29.3\% \quad (\textrm{Kerr}).
\end{eqnarray}
By the Penrose process, we can extract about
$29.3\%$  of the initial mass energy ${\cal M}_0({\cal A}_0,{\cal J}_0)$.
A
Schwarzschild black hole would be eventually left with its mass 
${\cal M}_f={\cal M}_{\rm irr}({\cal M}_0,{\cal J}_0)$.

Calculating these quantities using the formula in \cite{myers} 
for five dimensional black hole and black ring 
\cite{footnote},
we find
\begin{align}
 \label{eq:brirr}
{\cal M}_{\rm{irr}}=\frac{3(2\pi ^2{\cal A}^2)^{1/3}}{16\pi G}
=\frac{3\pi \lambda ^{1/3}(1+\lambda )^{1/3}
(\lambda-\nu)}{2^{5/3}G(1+\nu)^{4/3}(1-\nu )^{2/3}},
\end{align}
we find the rotational energy ${\cal E}_R={\cal M}-{\cal M}_{\rm irr}$ is 
\begin{eqnarray}
\varepsilon_R \equiv {\cal E}_{R}/{\cal M}=\left\{
\begin{array}{ll}
\displaystyle 1-\left(\frac{1-\nu}{1+\nu}\right)^{1/3}
\quad &(\textrm{a black hole}) \\
\label{eq:renergy}\\
\displaystyle 1-\left(\frac{\nu(1-\nu)}{2}\right)^{1/3}
\quad &(\textrm{a black ring}).
\end{array}
\right.
\end{eqnarray}

%======================= FIG. energy ============================%
\begin{figure}[h]
\begin{center}
\includegraphics[width=7cm]{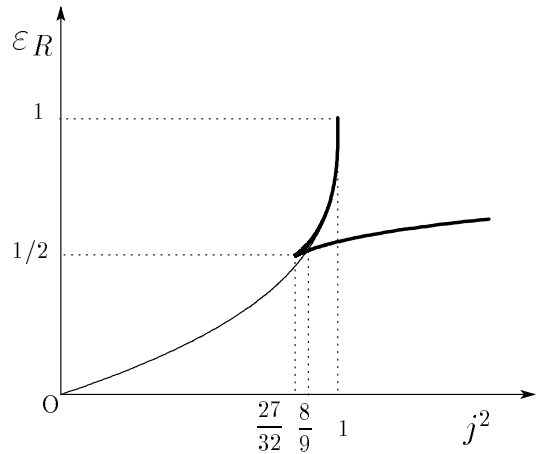}
\end{center}
\caption{Reduced rotational energy ($\varepsilon_R$) 
 is shown in terms of a reduced spin $j$. 
The fine line corresponds to a black hole solution, while the solid
 line to a ring solution. 
The overlapped point $(j^2,\varepsilon_R)=(1,1)$ is a naked
 singularity. The point $(0,0)$ represents 
the Tangherlini-Schwarzschild solution,
 so in this case we cannot extract energy from a black hole.}
\label{fig:energy}
\end{figure}
%----------------------------------------------------------------%
%
Fig.\ref{fig:energy} is the plot for 
$\varepsilon_R={\cal E}_R/{\cal M}$ against $j^2$. 
The rotational energy is monotonically increasing function of $j^2$ for MP
black holes with one angular momentum,
while it has a cusp at $j^2=27/32$ for a black ring solution. 
This point gives a lower bound of angular momentum of a black ring.
We also find a similar structure for the horizon area (entropy)-angular momentum relation
(see Fig. 3 in \cite{emparan}), in which we have also a cusp at the same point.
The higher entropy branch should be relatively
stable, while the lower entropy branch may be unstable, i.e.
it contains at least one unstable mode.
Note that the higher
 rotational energy branch corresponds to the lower entropy branch.
The appearance of such a cusp indicates change of stability.
 Such a behavior of stability
could be understood by a catastrophe theory \cite{torii}.

\section{Catastrophe theory and Stability}
\label{sec:catastrophe}
Catastrophe theory is a mathematical tool to explain a change of
stability in nature. 
In some phenomena, a state changes  discontinuously in spite of
a gradual change of the state parameters. The stability changes of
colored black holes, for example, are explained by a catastrophe theory
\cite{torii,tamaki}. The stability analysis via a catastrophe theory 
seems to have one-one
correspondence to the linear perturbations.

\begin{figure}[h]
\begin{center}
\includegraphics[width=8cm]{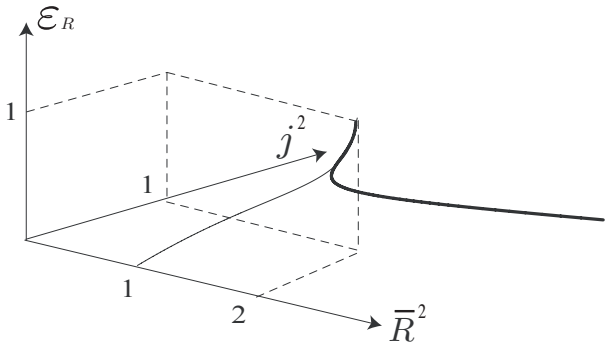}
\caption{Two smooth curves in the equilibrium space ${\cal V}=
(\bar{R}^2, j^2, \varepsilon _R)$:One (the fine line) corresponds to
a set of black hole solutions and the other  (the solid line)
to that of black rings.}
\label{fig:catastrophy}
\end{center}
\end{figure}

To see this more precisely, we introduce 
a potential function, which is a functional of  a control parameter
and a state variable.
We here assume that a control parameter is a radius of a black ring
$\bar{R}^2=3\pi R^2/2G{\cal M}$ 
(This can be driven by solving (\ref{eq:brmass}) for
$R$), a state variable is the reduced angular momentum $j^2$
and a potential function is the reduced rotational energy $\varepsilon _R$.
In Fig. \ref{fig:catastrophy}, we depict
 the equilibrium space ${\cal V}$.
We find that there are two smooth curves: One (the fine line) corresponds to
a set of black hole solutions and the other  (the solid line)
to that of black rings.
 The projection of the
equilibrium space onto the control plane 
(Fig.{\ref{fig:energy}}) is called a catastrophe map 
$\chi _V :{\cal V} \to \mathbb{R}^2$. At the point of the bifurcation 
$(\bar{R}^2, j^2, \varepsilon _R)= (25/12, 27/32, 1/2)$, 
the mapping  $\chi _V$ becomes singular. If such a singular point exist,
the stability changes there.

%========================== FIG Catastrophe =========================%
\begin{figure}[h]
\begin{center}
\includegraphics[width=8cm]{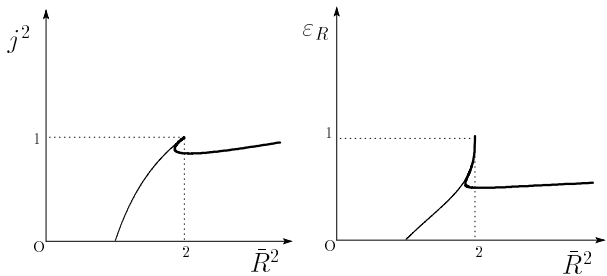}
\caption{The projection of the curve in the 
$(\bar{R}^2, j^2, \varepsilon _R)$ space onto the $(\bar{R}^2, j^2)$ plane 
and the 
$(\bar{R}^2,\varepsilon _R)$ plane.}
\label{fig:erjr}
\end{center}
\end{figure}
%=============================================================%

Since the irreducible mass is proportional to the two third powers of the
area (${\cal M}_{\rm irr}\propto {\cal A}_H^{2/3}$), 
the lower branch of the rotational energy has higher entropy,
which is thermodynamically stable configuration. 
When we compare the rotational energy of a black hole 
($\varepsilon_{R({\rm BH})}$)
and that of a black ring
in the lower branch ($\varepsilon_{R({\rm BR}^{(-)})}$),
we find that 
\begin{eqnarray}
&&
\varepsilon_{R({\rm BH})}<\varepsilon_{R({\rm BR}^{(-)})} 
~~~~{\rm for}~~ 
{27\over 32} <j^2 < {8\over 9} \\
&&
\varepsilon_{R({\rm BH})}>\varepsilon_{R({\rm BR}^{(-)})}
~~~~{\rm for}~~ 
{8\over 9} <j^2 < 1\,.
\end{eqnarray}
As a result, a black hole might be unstable for larger angular momentum, while
a black ring might be unstable for smaller angular momentum.
One may expect that there is a phase transition from a black ring
to a black hole when the object loses its angular momentum.
However, a catastrophe theory cannot predict it because
two curves in the equilibrium space are connected at a singular point 
($(\bar{R}^2, j^2, \varepsilon _R)= (2,1,1)$)
but not smoothly.
We do not predict any stability for
the equilibrium curve with such a singular point in a catastrophe theory.
Since two objects are topologically different,
classical perturbations will also not judge which spacetime is 
more stable.

We should also mention about  Gregory-Laflamme (GL) 
instability of a black ring \cite{gregory}.
When the angular momentum gets large, a ring becomes very thin, which phase
corresponds to the lower branch in Fig. \ref{fig:energy}.
Then we expect the similar instability to 
GL instability of a black string \cite{gregory},
because in the limit 
$R\to \infty ,\nu ,\lambda \to 0$ \cite{emparan2}, 
a black ring solution approaches a boosted black string solution
(see Appendix \ref{sec:app2}).
This GL instability seems to be different from
the above instability discussed in a catastrophe theory, which we shall
call the small 
entropy (SE) instability because the smaller horizon area (entropy)
branch would be unstable.
The upper branch of a black ring is
unstable in the sense of SE instability, while 
the lower branch (at least for large $j^2$) becomes unstable
in the sense of  GL instability.
Therefore,
it is not so clear what kind of stable configuration
is achieved for a black ring. 

\section{Concluding Remarks}
\label{sec:discussions}
 In this paper, we have discussed the energy extraction via the 
Penrose process from higher
 dimensional black holes and black rings. The result is that we can extract
 much more energy from higher dimensional black holes than from 4-dimensional 
Kerr black hole. Although we can gain at most about $20.7\%$ energy of incident
 particle in 4-dimensional 
Kerr space-time, the efficiency could be rather amplified in
 higher dimensions. In particular, if one rotation parameter vanishes,
the maximum efficiency becomes infinitely large because the angular momentum 
is not bounded from above.
We also apply a catastrophe theory to 
analyze  the stability of black rings.
Our analysis indicates a branch of black rings with higher
rotational energy is unstable, which should be a different
 type of instability from the Gregory-Laflamme's one.
The consequence might enable us to distinguish the
 4-dimensional Kerr black hole from higher dimensional 
ones by their energy extraction rates. 

However we have to be more careful.
The Penrose process tells us only 
possibility.
If we are interested in 
the microscopic process such as 
the Hawking evaporation of a black hole
or a superradiance,
we have to put some field in some background spacetime
and then quantize it.
The super-radiant mechanism in the MP black hole has been investigated in 
 \cite{frolov2, nomura}.
In the 5-dimensional black hole,
if one rotation parameter is much smaller (but not zero)
than the other,
superradiant modes play an important role, 
by which two rotation parameters turn
eventually to be almost equal \cite{nomura}.
For the black hole with the same rotation parameters,
it turns out that the thermal radiation is much
important than the superradiance even if the black hole is
maximally rotating.
This result would be consistent with our result,
i.e. the efficiency of energy extraction can be infinitely large
if one rotation parameter is very small, on the other hand
it is finite
for the case with the same rotation parameters.
 
As for a black ring,
so far we do not know so much.
Although the efficiency of energy extraction from 
a black ring can be infinitely large in the limit
 $\nu \to 1$ (\ref{eq:brefficiency}), the effective
 potential becomes larger and larger in this limit. 
When we quantized some field in this background,
the particles created by a quantum process
may not be able to escape to infinity.
If it is the case, the energy extraction via quantum process 
is not likely. 
To evaluate it,  we have to investigate
 the superradiance and Hawking radiation.
  The evaporation  mechanism and evolution
 of a black ring would tell us the real conclusion.
However, we have unfortunately 
not succeeded to separate the variables of geodesic
 motion and the Klein-Gordon equation in a black ring space-time.
What we can do so far is to analyze it in the limit of 
$j^2 \rightarrow \infty$ as ${\cal M}/R$ fixed,
which corresponds to a boosted black string spacetime.
We find that no superradiance occurs in this highly rotating spacetime.
The detail is given in Appendix \ref{sec:app2}. 
This result is again consistent with our result, i.e.
the efficiency of a black ring in the small-$\varepsilon_R$ branch  
decreases with the angular momentum
$j$. However, the Penrose process is still possible because
the efficiency reaches some finite constant even when $j\rightarrow \infty$

 The stability of the
 black  ring as well as higher dimensional rotating black hole,
 has not yet been established. Only the stability of
 Schwarzschild black hole is proved in higher dimensions
 as well \cite{kodama}.  An interesting
 thermodynamics approach to stability shows that 
 they are unstable in
 specific limit \cite{emparan2,emparan3}. A black ring
 approaches a boosted black string and a rotating black hole with large
 angular momentum in $D\geq 6$ approaches a black membrane. Black
 strings and membranes are shown to be unstable \cite{gregory}. 
 It would be interesting to perform a stability analysis by the metric
 perturbations and compare the result with a thermodynamics viewpoint. 

 As we demonstrated in section \ref{sec:im}, the appearance of cusp in
 Fig.\ref{fig:energy} indicates that there is a stability change at that point
via  a catastrophe theory. This shows that a black ring has another 
unstable mode which is different from the GL instability. 
It may be interesting to clarify what kind of instabilities exist
in a black ring spacetime.

\section{Acknowledgements}
We are grateful to T. Torii for  fruitful discussions.
This work was partially supported by the Grant-in-Aid Scientific Reseach
Fund of the MEXT (No.14540281), The 21st Century 
COE Program (Holistic Research and Education Center for Physics 
Self-organization Systems) at Waseda University and the Waseda University Grant for
Special Research Projects .

\appendix

\section{The effective potential of a test particle in a black ring spacetime}
\label{sec:app1}
We discuss here about a particle motion in a black ring spacetime.
We assume that a particle moves on the ``outside''
 equatorial plane, i.e.
$x=-1$.
Because $y$ coordinate is singular on the ergosurface,
we introduce another coordinate $z$, which is defined by $z=-\tanh ^{-1}y^{-1}$.
Using this coordinate, we write down the effective potential 
 on the plane $x=-1$ for a particle with the energy $E$
and angular momentum $L$,
 which is obtained from $p^\mu
p_\mu =-m^2$ ($m^2=0$ for a null particle), as
\begin{align}
 \left(\frac{dz}{d\tau }\right)^2 &=\varepsilon (A E^2-2B
 E+C)\nonumber \\
&= \varepsilon A (E-V_+)(E-V_-) ,
\end{align}
where
\begin{align}
V_{\pm }&=\frac{B \pm \sqrt{B^2 -AB}}{A} ,
\end{align}
and
\begin{align}
\varepsilon&= \frac{1-\tanh z}{R^2(1+\lambda)^2
(\tanh z+\lambda )(1+\tanh z)^2}\\
A &=\lambda \nu (\tanh ^2 z-3\tanh z+4)+(\tanh z+\lambda +\nu),\\
B &=\frac{L}{R}\sqrt{\frac{\lambda
 \nu}{1+\lambda}}(1+\nu)(1-\tanh z)^2 ,\\
C &=\frac{L^2}{R^2}(1+\nu)^2(1+\lambda)\tanh z (\tanh
 z-1)\nonumber \\
& \quad-m^2(1+\lambda)(1+\tanh z)(\tanh z+\nu).
\end{align}

\begin{figure}[h]
\begin{center}
\includegraphics[width=7cm]{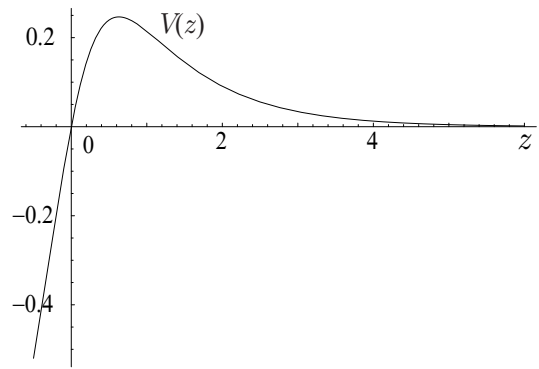}
\end{center}
\caption{The effective potential for a black ring. Here we used a new
 coordinate $z=-\tanh ^{-1}y^{-1}$. The horizon corresponds to $z=-\tanh^{-1}
 \nu $ and the infinity to $z=\infty $. The plot is for $\nu =1/2$ and
 the vertical line is normalized by $L/R$.}
\label{fig:ring}
\end{figure}

In Fig.\ref{fig:ring}, we depicts the effective potential for a null
particle against $z=-\tanh ^{-1}y^{-1}$. 
In this coordinates, asymptotic infinity is at $z\to +\infty $. 
The negative energy states
exist in the ergoregion 
$-\tanh^{-1}\nu <z<0$. 

%----------------------------------------%
%----------------------------------------%

\section{No superradiance of a black ring in the limit of 
$j^2\rightarrow\infty$}
\label{sec:app2}
We show that superradiance does not occur from a very large black ring.
A black ring solution approaches a boosted black string solution 
in the limit of $j^2\rightarrow \infty$ as ${\cal M}/R$ is 
fixed \cite{emparan2}. In this limit,
\begin{equation}
R\rightarrow \infty, \qquad \lambda ,\nu \rightarrow 0\,,
\label{boost_limit}
\end{equation}
with keeping $R\lambda$ and $R\nu$ constant.
Introducing new coordinates $r,\theta,\omega$ and 
parameters $r_{H},\sigma$ as
\begin{eqnarray}
 \nonumber
&&r=-RF(y)/y,\quad \cos \theta =x, \quad \varpi =R\psi \,,\\
&&R\lambda =r_{H}\cosh ^{2}\sigma ,
\qquad R\nu =r_{H} \sinh ^{2}\sigma \,.
\label{eq:string}
\end{eqnarray}
and taking the above limit (\ref{boost_limit})
 of a  black ring solution
 $(\ref{eq:blackring})$, we obtain a boosted black string solution  
%----------------------boosted black string--------------------%
\begin{align}
\label{eq:boosted}
ds^{2}=-\bar{f}\left(dt-\frac{r_{H}\sinh2\sigma}{2r\bar{f}}d\varpi\right)^{2}
+\frac{f}{\bar{f}}d\varpi ^{2}+\frac{1}{f}dr^{2}+r^{2}d\Omega_{2} ^{2} 
\end{align}
with
\begin{eqnarray}
f&\equiv& 1-\frac{r_{H}}{r}, \qquad \bar{f}\equiv 1-\frac{r_{H}\cosh
 ^{2}\sigma}{r}\,,\\
d\Omega_{2} ^{2}&=&d\theta^2+\sin^2\theta d\phi^2,
\end{eqnarray}
where  $\sigma$ is a boost parameter.
$\sigma =0$ corresponds to a (static) black string solution.
The horizon exists at  $r=r_{H}$.
The Killing vector $\xi =\partial _t$ becomes
null at the horizon.
The ergoregion also exists ($r_{H}<r<r_{H}\cosh^{2}\sigma $).

We consider a massless free scalar field $\Phi $ in 
a boosted black string spacetime (\ref{eq:boosted}).
The basic equation for $\Phi$ is 
%-------------------------Klein-Gordon equation------------------------%
\begin{equation}
\Box \Phi =\frac{1}{\sqrt{-g}}\partial _{\mu }(\sqrt{-g}g^{\mu \nu }\partial
 _{\nu }\Phi )=0 ,
\label{eq:klein}
\end{equation}
which is
separable in the spacetime (\ref{eq:boosted}). Here we set
\begin{equation}
\Phi = \frac{F(r)}{r}Y_{lm}(\theta ,\phi )e^{ikz}e^{-i\omega t}
\end{equation}
where $Y_{lm}$ denotes spherical harmonics.
Then, we find the radial equation of (\ref{eq:klein}) as
%------------------------radial equation---------------------------%
\begin{eqnarray}
\label{eq:radial}
\frac{d^{2}F}{dr_{*}^2}+V(r)F=0 ,
\end{eqnarray}
with
\begin{eqnarray}
V(r)&= &f\left(\frac{\omega ^2}
{\bar{f}}-\frac{f'}{r}-\frac{l(l+1)}{r^2}\right)
\nonumber
\\
&&~~~~~
-\bar{f}\left(k+\frac{r_{H}\sinh 2\sigma}{2r\bar{f}}\omega \right)^{2},
\end{eqnarray}
where $r_{*}$ is the tortoise coordinate
%--------------------tortoise coordinate----------------------------%
\begin{eqnarray}
\frac{dr_{*}}{dr}&=&1/f ,\nonumber\\
\enspace r_{*}&=&r+r_{H}\ln |r-r_H|.
\end{eqnarray}
%------------------------potential-----------------------%
The potential function $V$ asymptotically approaches some constants
as
\begin{eqnarray}
V(r) \sim \left\{
\begin{array}{ll}
 \omega_\infty^2 ~~~~
 &(r \rightarrow \infty)\,\,\\
 \omega_H^{2} ~~~~ &(r \rightarrow r_H)
\,\,,
\end{array}
\right.
\end{eqnarray}
where 
\begin{eqnarray}
{\omega_\infty}&=&\left(\omega ^{2}-k^2\right)^{1/2}  \,,\\
{\omega_H}&=&\cosh \sigma (\omega -k\tanh \sigma )    \,.
\end{eqnarray}

~\\
To discuss the scattering of the scalar wave in this spacetime,
we consider an incoming wave with a unit amplitude  $F_{\omega lmk}$,
 which asymptotic forms are given by 
%---------------------- in and out -----------------------%
\begin{eqnarray}
F_{\omega lmk} \sim \left\{
\begin{array}{ll}
 e^{-i{\omega_\infty}r_{*}}+A_{\omega mlk}e^{i{\omega_\infty }
 r_{*}}\,\, &(r\to \infty )\\
 B_{\omega lmk}e^{-i{\omega_H}r_{*}} \,\,
 &(r\to r_H)
\end{array}
\right.
\end{eqnarray}
Since the Wronskian is constant with respect to $r_{*}$ for the solutions
of (\ref{eq:radial}) and their complex conjugates, we find
\begin{eqnarray}
1-|A_{\omega lmk}|^{2}=\frac{\omega_H}{\omega_\infty}
|B_{\omega lmk}|^{2} 
\end{eqnarray}
%--------------------------superradiance mode-------------------------%
Since the wave we are discussing is travelling at infinity,
 we have a constraint that
$\omega_\infty>0$, which gives $\omega >k$.
Then the condition for the superradiance ($|A_{\omega lmk}|>1$) is 
\begin{eqnarray}
{\omega_H}<0 ~~ \Longleftrightarrow ~~ \omega <k\tanh \sigma  \,,
\end{eqnarray}
which leads to $\omega < k$.
This is not consistent with our previous condition for $\omega$.
Hence we come to the conclusion that 
superradiance does not occur for a
boosted black string.

\begin{figure}[h]
\begin{center}
\includegraphics[width=7cm]{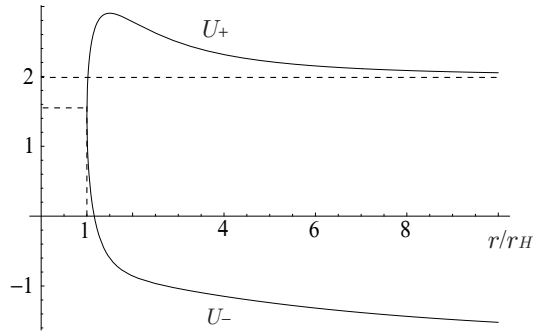}
\end{center}
\caption{The effective potential $U_\pm$ for a boosted black string.
We set $L_\varpi =2, L_\phi =6, \sigma =1$.}
\label{fig:SR_BR}
\end{figure}

The superradiance occurs by the difference between
minimum energy of a particle at
the event horizon and that at infinity.
In the Kerr black hole,  due to the inertia frame dragging,
the minimum energy at the horizon is
raised by the rotation of a black hole 
from 0 to $m\Omega _{H}$, where $m$ is a magnetic quantum number.
Hence, if $0<\omega<m\Omega _{H}$, we have superradiance.
In the present case, the minimum energy of a particle at the horizon
cannot be raised enough high compared with that at infinity 
(see Fig.\ref{fig:SR_BR}). 

%------------------------ radial motion -------------------------%
To see this more precisely, we write down the radial  equation
 for a massless particle, i.e. the
equation  $p_\mu p^\mu =0$ leads to
\begin{align}
\dot{r}^2 &=\tilde{f} E^2
+\frac{r_H\sinh 2\sigma }{r}EL_\varpi
-\bar{f}L_\varpi  ^2+\frac{f}{r^2}L_\phi ^2 \\
&=\tilde{f}(E-U_+)(E-U_-),
\end{align}
with 
\begin{align}
\tilde{f}&\equiv 1+\frac{r_H\sinh ^2 \sigma }{r}, \\
U_\pm &=\frac{1}{\tilde{f}}\left[\frac{r_H\sinh 2\sigma }{2r}L_\varpi
\right.
\nonumber \\
&~~~~~~~
\left.
\pm \sqrt{\left(\frac{r_H\sinh 2\sigma }{2r}L_\varpi \right)^2
+\tilde{f}\left(\bar{f}L_\varpi ^2+\frac{f}{r^2}L_\phi ^2 \right)}\right],
\end{align}
where $E, L_\varpi $, and $L_\phi$ represent the energy, the 
$\varpi $- and $\phi$- component of the angular momentum
of a particle, respectively.
The effective potential $U_\pm $ asymptotes the value
\begin{align}
U_\pm \sim \left\{
\begin{array}{ll}
\pm L_\varpi   &(r\to \infty )\\
L_\varpi \tanh \sigma  &(r\to r_H)\,,
\end{array}
\right.
\end{align}
which gives
the minimum energy of a particle at infinity 
or that at horizon.
Fig. \ref{fig:SR_BR} shows the typical behavior of the 
effective potential $U_\pm (r)$
 for a particle.
$U_+ (r)$ gives 
the minimum energy of the particle at $r$.

Although there is no superradiance
in a boosted black string spacetime,
the Penrose process is still possible.
In fact, the maximum efficiency 
for a black ring is finite ($\eta_{\rm max}=(\sqrt{2}-1)/2$) 
even in the limit of
$j^2\rightarrow \infty$.

%====================== FIG. super ==========================%
%=============================================================%

%============================================================%
%====================   References  =========================%
%============================================================%


\begin{thebibliography}{99}
\bibitem{footnote}
Since a black ring must be rotating (there is a lower bound of $j$)\cite{weyl},
it is not trivial to divide the mass energy into 
its irreducible mass and the rotational energy.
 Here we assume that the same formula for irreducible mass is applicable
 for a black ring as well.
\end{thebibliography}
\end{document}